\documentclass[preprint,prb,superscriptaddress,noshowpacs,floatfix]{revtex4}
\usepackage{bm}
\usepackage{graphicx}

\def\bea {\begin{eqnarray}}
\def\eea {\end{eqnarray}}
\def\be {\begin{equation}}
\def\ee {\end{equation}}

\def\pp{^\prime}

\def\fact{{1 \over N \beta}}
\def\g2q{g({\bf q},i\nu_n)}

\def\bea {\begin{eqnarray}}
\def\eea {\end{eqnarray}}
\def\be {\begin{equation}}
\def\ee {\end{equation}}

\def\pp{^\prime}

\def\fact{{1 \over N \beta}}
\def\g2q{g({\bf q},i\nu_n)}

\def\bea {\begin{eqnarray}}
\def\eea {\end{eqnarray}}
\def\be {\begin{equation}}
\def\ee {\end{equation}}
\def\pp{^\prime}
\def\up{\uparrow}

\def\co{c_{{\bf m}_1\uparrow} (\tau_1)}

\def\cop{c^\dagger_{{\bf m}_1\pp\uparrow} (\tau_1\pp)}

\def\g2q{g({\bf q},i\nu_n)}

\def\fact{{1 \over N \beta}} 

\begin{document}
\title{Demonstration of a robust pseudogap \\
in a three-dimensional correlated electronic system} 
\author{R.J. Gooding}
\affiliation{Dept of Physics, Queen's University, Kingston ON K7L 3N6}
\author{F. Marsiglio}
\affiliation{Dept of Physics, University of Alberta, Edmonton AB T6G 2J1}
\author{S. Verga}
\affiliation{Dept of Physics, University of Alberta, Edmonton AB T6G 2J1}
\author{K.S.D. Beach}
\affiliation{Department of Physics, MIT, Cambridge, MA 02169}
\date{\today}
\newpage
\begin{abstract}
We outline a partial-fractions decomposition method for determining the
one-particle spectral function and single-particle density of states 
of a correlated electronic system on a finite lattice 
in the non self-consistent T-matrix approximation {\em to arbitrary numerical
accuracy}, and demonstrate the application of these 
ideas to the attractive Hubbard model. We then demonstrate the effectiveness of a 
finite-size scaling ansatz which allows for the extraction of quantities of interest 
in the thermodynamic limit from this method. In this approximation, in one or two 
dimensions, for any finite lattice or in the thermodynamic limit, a pseudogap is 
present and its energy
diverges as $T_c$ is approached from above; this is an unphysical manifestation of using 
an approximation that predicts a spurious phase transition in one or two dimensions. However, 
in three dimensions one expects the transition predicted by this approximation to represent 
a true continuous phase transition, and in the thermodynamic limit any pseudogap predicted
by this formulation will remain finite. We have applied our method to the attractive Hubbard 
model on a three-dimensional simple cubic lattice, and find, similar to previous work, that 
for intermediate coupling 
a prominent pseudogap is found in the single-particle density of states, and this
gap persists over a large temperature range. In addition, we
also show that for weak coupling (an on-site Hubbard energy equal to one quarter the bandwidth)
a pseudogap is also present. The pseudogap energy at the transition temperature
is almost a factor of three larger than the $T$=0 BCS gap for intermediate coupling, whereas
for weak coupling the pseudogap and $T$=0 BCS gap energies are essentially equal. 
These results show that a pseudogap due to superconducting fluctuations occurs in three 
dimensions even in the weak-coupling limit. 
\end{abstract}
\maketitle

\section{INTRODUCTION}
\label{sec:int}

The theory of Bardeen, Cooper, and Schrieffer (BCS) \cite{bardeen57} 
has been very successful in explaining a wide variety of phenomena 
associated with superconductivity. This is all the more remarkable
since it is based on the independent quasi-particle approximation.
Since very early on corrections to this simple picture have been
investigated to explore what effect, if any, the breakdown of 
the quasi-particle approximation would have on various superconducting
properties. One body of work in this category is the investigation
of collective excitations, beginning with Anderson,\cite{anderson58} 
Bogoliubov {\it et al.} \cite{bogoliubov59}, Rickayzen \cite{rickayzen59}
and others.\cite{bardasis61} Much of this work was directed towards
a proper gauge-invariant description of the Meissner effect.

A separate line of work started with Migdal \cite{migdal58} and
Eliashberg,\cite{eliashberg60} and investigated how the 
coupling of electrons to dynamical phonons would affect the 
quasiparticle properties of the electrons, including the 
transition temperature.

While this and other work \cite{parks69} focussed on properties in
the superconducting state, a different line of enquiry 
concerned pairing fluctuations above $T_c$, as these could possibly 
cause deviations from the mean field picture provided by Fermi Liquid/BCS theory.
Within BCS theory superconductivity arises through the formation
of an order parameter, which has both an amplitude and a phase.
\cite{landau50} Fluctuations can impact both the amplitude and the
phase. At the present time phase fluctuations have been suggested to be
relevant to the high-$T_c$ problem. In particular, a school of thought
\cite{alexandrov95,feinberg90,emin93,ranninger95} suggests that electrons have a pair 
amplitude already above $T_c$, i.e. they exist as `pre-formed' pairs.
The superconducting transition is then simply the temperature at which
the pairs acquire a common phase.\cite{emery95} The extreme case
of this scenario is where two electrons already form a 
well-defined boson at the critical (i.e. phase-forming) temperature; 
this temperature is then the Bose condensation temperature for 
the system. A comprehensive review of such ideas has been provided
by Randeria.\cite{randeria95} 

These so-called pairing fluctuations would generally lead
to a non-zero amplitude for pairing above $T_c$, with long-ranged
superconducting order destroyed by fluctuations. One school of thought
(to which we subscribe) suggests that some of the anomalous normal state
properties observed in the high $T_c$ systems, in particular the pseudogap,
are due to pairing fluctuations, and a goal in this paper is the examination 
of such fluctuations in different dimensionalities using one particular 
controlled approximation scheme.

Specifically, one technical framework with which one can hope to understand
this physics is the so-called T-matrix approach. Within such a scheme,
a pairing susceptibility can be defined;\cite{allen79} in the usual 
case,\cite{ambegaokar69} this pairing susceptibility diverges at $T_c$, 
the latter defined by BCS theory. Finding this temperature constitutes an alternate 
means of defining the transition temperature, and is generally known as 
the Thouless criterion.\cite{thouless60} 
However, the existence of a diverging susceptibility (as one approaches the BCS
critical temperature from above) should impact the single electron properties
that determine the critical temperature in the first place. The T-matrix
approximation accomplishes this through a renormalization of the potential an
electron experiences. However, a limitation of this approach is that it can
only be justified as a low density expansion of the two-particle Green's function.
\cite{kanamori63,fetter71}

Several many-body frameworks within which the T-matrix approximation can
be cast have been discussed by Baym and
Kadanoff and coworkers \cite{kadanoff61,baym61,baym62} many years ago.
The particular choice of framework unfortunately influences the results one
obtains when calculations for a particular model system are actually performed.
For example, in Ref.,\cite{kadanoff61} a particular prescription for 
the single electron self-energy is derived through an equation-of-motion method.
The resulting expression requires the use of two `levels' of single electron Green functions,
one bare, the other renormalized; that is, the latter contains a 
self-energy, and, consequently, such calculations must proceed self-consistently.
Nonetheless, already in this work\cite{kadanoff61} (see footnote 13 and the 
discussion at the end of section 3) it was recognized that certain features of this 
approximation were not acceptable.

Several years later the issue was debated some more,
\cite{schmid70,marcelja70,patton71} until what one could refer to as the
`Patton fix' emerged as the allegedly correct way to handle the difficulties with
the Kadanoff-Martin prescription. In this adjustment, one of the 
self-consistent single electron Green functions is simply replaced with
its non-self-consistent bare counterpart. This corrects some problems, 
but the adjustment is completely {\it ad hoc}, and has been called
``non $\phi$ derivable".\cite{sharapov01} Such an approach 
has been championed recently by Levin and coworkers,\cite{janko97}
and has found some agreement with experiments on the high $T_c$ cuprates.

Earlier, Baym \cite{baym62} had formulated a theory of single electron and 
two particle properties based on functional derivatives of a free energy functional, 
a procedure which guaranteed that certain conservation laws would be satisfied.
This formulation is known as a conserving approximation, and its
extension to lattice electrons is known as the FLuctuation Exchange approXimation
(FLEX) \cite{bickers89}. When only the particle-particle channel is
retained, the FLEX becomes a T-matrix theory with fully self-consistent
single electron Green functions. 
Such a formulation would appear to be the most accurate of the T-matrix
approximations, since it includes the most number of diagrams. However,
this claim is, in our opinion, unfounded. The possibility that vertex
approximations (omitted in all T-matrix theories discussed here) partially
cancel some of the self-energy contributions retained in the more fully
self-consistent T-matrix approximations appears likely.\cite{verga04}
That is, including more channels and making everything self-consistent
is not the panacea it might seem to be --- this point has been emphasized
in the alternative formulation of this problem by Vilk and Tremblay.\cite{vilk97}

It is probably fair to say that `modern' interest, and a viable theoretical
approach, on the influence of pair fluctuations on the superconducting transition 
was initiated by Leggett.\cite{leggett80} This was later followed by an important paper 
by Nozi\'eres and Schmitt-Rink,\cite{nozieres85} wherein
they solved for the zero temperature properties of an electron gas with
attractive interactions, for all coupling strengths. They showed that the
BCS approximation gave the correct limiting behaviour in the strong coupling
limit, where, in the lattice model at least, it was clear that the model
was best described by tightly bound pairs of electrons. In this limit, it
became clear that the superconducting transition was properly described
by a Bose condensation of electron pairs. An important technical point
raised in these two papers was that both the BCS gap equation and number equation 
were required, and had to be solved self-consistently.

This work was further extended to two dimensions in the context of the 
high $T_c$ cuprates, by Schmitt-Rink, Varma, and Ruckenstein, in 
Ref. \cite{schmittrink89} They made the point that, for a specific (and
fixed) electron  density, as the temperature was lowered the chemical potential would 
self-consistently adjust so as to avoid the Thouless instability. (Note
the necessity of adhering to the lesson from Ref.,\cite{leggett80}
that being that one must solve both for the normal state properties and the
electronic density simultaneously.) In fact, all electron densities would result in a
situation at $T=0$ in which the Fermi surface no longer exists; the electrons
are paired in bose-like bound states, and the chemical potential is simply
half of the single pair bound state energy. This result violates
Luttinger's theorem.\cite{luttinger60} Improvements were made by
Serene,\cite{serene89} Tokumitu {\it et al.} \cite{tokumitu93},
and later, for lattice fermions, by three of the present 
authors,\cite{beach01} but the physics obtained in Ref. \cite{schmittrink89} 
remains. In fact, for lattice fermions the results are more physical, 
as the Thouless curve becomes a locus of points in the $T- \mu$ plane 
on which the electron density is identically equal to unity\cite{beach01} (as
opposed to a diverging quantity). Again, the simultaneous solution of the 
superconducting transition and the equation for the electronic density is required.

Since then a significant number of calculations have appeared
on the T-matrix problems; see, for example, Refs. 
\cite{fresard92,haussmann93,kagan98,kyung98,gusynin99,engelbrecht00,rohe01,yanase01,quintanilla01}
In addition, arguments against such formulations and in favour of alternate
frameworks requiring vertex corrections, have also appeared.\cite{vilk97,pieri98,allen01,kyung01}
This body of work represents a significant advance in the study of the effect of fluctuations
at temperatures above the superconducting transition. However, much of the `mathematical
contortions' that are invoked can be traced to the difficulty of studying two-dimensional
systems in which the Mermin-Wagner theorem precludes the existence of true long-ranged
order (in the systems studied here).\cite{KT} Self consistency, and other formulations,
aim to recover this result from any Thouless criterion approach (by which we imply theory
attempting to identify $T_c$ through a diverging pair susceptibility). 

A simple remedy to avoid the Mermin-Wagner theorem is to study a three dimensional system.
Therefore, in this paper we focus on a purely electronic Hamiltonian that yields a nonzero
superconducting transition temperature, the attractive Hubbard model for 
a simple cubic lattice. Further, in three dimensions we can avoid the necessity of using 
a self-consistent theory, especially for weak coupling, and this is a simplification which 
will allow for us to solve for the self energy of a finite lattice, using the method outlined 
in the following sections of this paper, to essentially any numerical accuracy that we require.

In the above-described work, the appearance of a pseudogap, an experimentally observed 
phenomenon in which the single-particle density of states is suppressed in the form of an 
incompletely formed 
superconducting gap,\cite{statt} is predicted. The literature on this question is now 
substantial --- for a review of recent ideas see, for example, Ref. \cite{sharapov01}.
A recurring theme in some of this work (see, \textit {e.g.} Ref.\cite{kyung01}) 
is that the pseudogap found in the high $T_c$ cuprates can be 
associated with the two dimensional nature of the CuO planes. In terms of fluctuations, and in
its simplest form, this is simply the statement that one expects enhanced fluctuations, towards the
low-temperature phase, the lower the effective dimensionality of the system. The analysis that we 
present in this paper suggests that two dimensions may, in fact, not be necessary, because a 
pseudogap is found in the three dimensional attractive Hubbard model at weak coupling. 
For example, even at a temperature of a few per cent above the superconducting transition, in the
three-dimensional attractive Hubbard model at half filling, clear evidence of a single-particle
density of states pseudogap is found, and the magnitude of the gap is comparable to the
BCS gap that would be found at $T=0$.

We note that the number of papers on the 3d pseudogap is very small, compared to that for the 2d
problem, so comparisons with other work are difficult. Instead, here we shall simply address
the question, should one expect a pseudogap in the 3d system with purely electronic attractive
interactions?

\section{FORMULATION}
\label{sec:form}

\subsection{Model}
\label{subsec:model}

We consider lattice fermions and the 
attractive Hubbard model; we restrict our attention
to a kinetic energy that includes nearest neighbour hopping only. 
Working in the grand canonical ensemble thus leads us to consider the so-called
Grand Hamiltonian ${\cal K}$ defined by 
\be
\label{hubbard_hamiltonian}
{\cal K} = -t~\sum_{\langle {\bf m},{\bf m^\prime} \rangle \atop \sigma} 
(c_{{\bf m}\sigma}^\dagger c_{{\bf m^\prime},\sigma} 
+c_{{\bf m^\prime} \sigma}^\dagger c_{{\bf m},\sigma}) 
- \mu~\sum_{{\bf m},\sigma} 
n_{{\bf m}\sigma} 
-|U|~\sum_{{\bf m}} n_{{\bf m}\up}n_{{\bf m}\downarrow}~.
\ee
In this equation $t$ is the single particle hopping matrix element for
nearest neighbour sites, $\mu$ is the chemical potential, $|U|$ is the
on-site attractive potential, ${\bf m}$ and ${\bf m^\prime}$ label the 
lattice sites, while $\langle {\bf m},{\bf m^\prime} \rangle$ denotes 
nearest neighbour sites (which are counted only once in the kinetic
energy sum), and $\sigma$ is the electron spin. The electron creation 
(annihilation) 
and number operators with spin $\sigma$ at lattice site ${\bf m}$ are given by
$c_{{\bf m}\sigma}^\dagger$ ($c_{{\bf m}\sigma}$) and $n_{{\bf m}\sigma}$, 
respectively. We restrict ourselves to one-dimensional chains, and 
three-dimensional simple cubic lattices, and employ periodic boundary 
conditions.  

\subsection{One and Two Particle Propagators}
\label{subsec:form}
 
Firstly, we define the required thermal Green's 
functions. The single electron Green's function\cite{rickayzen80} monitors 
the history of an electron injected at (imaginary) time $\tau_1^\prime$ and 
lattice site ${\bf m}_1^\prime$, and 
removed at time $\tau_1$ and site ${\bf m}_1$, and is defined by
\be
\label{single}
G_\up({\bf m}_1,\tau_1; {\bf m}_1\pp \tau_1\pp) = (-)\langle {\rm T}_\tau [ \,
\co \cop ] \rangle \equiv G_\up(11\pp) ~~\equiv G(1-1\pp) ~.
\ee
We use standard abbreviating notation, $1 \equiv ({\bf m}_1,\tau_1)$,
and note that the last result of Eq.~(\ref{single}) follows from the translational 
periodicity, the time invariance, and spin isotropy of our lattice and 
Hamiltonian.  Similarly, the two-electron propagator is given by 
\bea
\label{two-general}
&&{\cal G}_{\uparrow \downarrow}({\bf m}_1,\tau_1; {\bf m}_2, \tau_2; 
{\bf m}_1\pp, 
\tau_1\pp; {\bf m}_2\pp, \tau_2\pp) ~\equiv~ {\cal G}(12;1\pp2\pp) \nonumber \\
&&=~\langle {\rm T}_\tau \
[ c_{{\bf m}_1\up} (\tau_1)               c_{{\bf m}_2\downarrow} (\tau_2)
c^\dagger_{{\bf m}_2\pp\downarrow} (\tau_2\pp) c^\dagger_{{\bf m}_1\pp\up} 
(\tau_1\pp) ]
\ \rangle~. 
\eea 
(In this paper we will need to only consider the 
${\cal G}_{\uparrow \downarrow}$ function,
{\em viz.} the singlet channel, and thus spin indices of $\cal{G}$ will not be
shown.) This latter function's transform in 
momentum/energy space can be simplified by incorporating translational 
periodicity and temporal invariance, and we introduce the notation $Q \equiv ({\bf Q},i\nu_Q)$ and 
$Q1 = {\bf Q}\cdot{\bf m}_1~+~i\nu_Q\tau_1$. Then we have
\be
{\cal G} (12;1\pp2\pp)~=~\frac{1}{{(\beta N)}^3}~\sum_{Qqq\pp}
e^{i\frac{Q}{2}((1+2)-(1\pp+2\pp))}
~e^{iq (1-2)}~e^{iq\pp(1\pp-2\pp)}~
{\cal G} (Q;qq\pp)
\label{two-general-FT}
\ee
where $N$ is the number of lattice sites, $Q$ is the centre-of-mass generalized
momentum of either the outgoing or incoming pair, and $q$ ($q\pp$) is the relative 
momentum of the outgoing (incoming) pair of electrons, respectively. All 
wavevector summations span the first Brillouin zone, and Matsubara sums go 
over all integers. 

From this function we extract the quantity that is our principal concern, 
{\textit viz.} the pair propagator ${\cal G}^{pair}$ 
which can be included in this set of equations in the following two ways:
\bea
\label{pair-FT1} 
{\cal G}^{pair} (Q)~&=&~\frac{1}{{(\beta N)}^2}~\sum_{qq\pp}~ 
{\cal G} (Q;qq\pp) \\
{\cal G}(11^+;1\pp 1^{\prime +})~&=&~\frac{1}{(\beta N)}~\sum_{Q}~ e^{iQ(1-1\pp)}
{\cal G}^{pair}(Q) ~.
\label{pair-FT2}
\eea
An examination of Eqs.~(\ref{two-general-FT}-\ref{pair-FT2}) shows that the 
pair propagator
is equal to the probability of adding a pair of (opposite spin) electrons with 
one particular centre-of-mass momentum $Q$ to {\em one} point in space, and 
then removing them as a pair at some other point in space (the centre-of-mass 
momentum $Q$ is conserved at all collisions that occur between these points 
in time, and one must sum over all possible relative momenta, $q$ and $q\pp$).

The pair propagator's connection to superconductivity can be made apparent 
by introducing the pair annihilation operator via
\be
\Delta_{\bf Q} ~=~ \frac{1}{\sqrt N} \sum_{\bf q}~
c_{\frac{\bf Q}{2} + {\bf q}\up}
c_{\frac{\bf Q}{2} - {\bf q}\downarrow}
\label{Delta-pair}
\ee
and then noting, by Eq.~(\ref{pair-FT1}), that the pair propagator can be 
expressed as
\be
{\cal G}^{pair}(Q)~=~\int_0^\beta~e^{i\nu_Q \tau} \langle T_\tau [ \Delta_{\bf Q} (\tau)
\Delta_{\bf Q}^\dagger (0) ] \rangle~d\tau~~.
\label{Delta_2_pair}
\ee
This equation most clearly allows one to understand the association of the 
pair propagator with the so-called  pair susceptibility.\cite{allen79}

Finally, it is instructive to extrapolate these relations to real times --- 
one sees that one other way to interpret the behaviour of the pair propagator 
is that its analytic continuation from imaginary frequencies to the complex 
plane allows one to evaluate the retarded pair propagator, in real time. As 
the discussion of Ambegaokar\cite{ambegaokar69} makes apparent, the poles 
of the pair propagator correspond to the relaxation times of a pair described by 
the retarded Green's function (also see Ref. \cite{kadanoff61}).  
Most of the poles correspond to continuum states comprised simply of two single 
(quasi-) electron states. Two poles are exceptions; initially these occur on 
the real axis. However, at sufficiently low temperature these poles form 
conjugate pairs on the imaginary axis. One of these poles signals an 
exponentially unstable ({\em viz.} growing) paired state implying that the 
normal state has now become unstable with respect to some new state whose
stability is presumably related to these paired electrons. It is clear 
that, using this approach, one cannot proceed to temperatures below 
the critical temperature $T_c$. On the other hand, studying the pair 
propagator above this critical temperature is an ideal way to study pair 
fluctuations, as well as other properties of the fully interacting normal 
state. This formalism for identifying $T_c$ from above has come to be known 
as the Thouless criterion.\cite{thouless60}

\subsection{Non Self-Consistent T-Matrix Approximation}

The above discussion motivates the analysis of the pair propagator in the 
normal phase. In order to find this function for the AHM we need to invoke 
a number of approximations. To be specific, we employ the ladder approximation 
to the Bethe-Salpeter equation for the reduced vertex. The derivations associated 
with this approximation, which leads to the T-matrix approximation for the single-particle
propagator, are well known. It has been shown that this 
approximation is justified (see, {\it e.g.}, Ref. \cite{fetter71}) in the 
limit of vanishing electron (or hole) densities.

The physical content of this approximation can be made most apparent if we 
begin with the equation for the 2-particle Green's function expressed in terms 
of the effective 2-particle interaction, which is given by (again, suppressing 
spin indices and sums over these indices)
\bea
&&{\cal G} (Q;q,q\pp) = \beta N \delta(q-q\pp) 
G^0 (\frac{Q}{2} + q) G^0(\frac{Q}{2} - q) \nonumber \\ &-&G^0 (\frac{Q}{2} + q) 
G^0 (\frac{Q}{2} - q)
\Gamma(Q;q,q\pp) G^0(\frac{Q}{2} + q\pp) G^0(\frac{Q}{2} - q\pp)
\label{G2_LA_eqn}
\eea
where $G^0$ is the noninteracting single-electron Green's function.
The ladder sum for the effective interaction of the AHM, in a 
non self-consistent (NSC) calculation, is given by
\be
\label{Gamma_LA_eqn}
\Gamma (Q;q,q\pp) ~=~-|U|  + |U|~{\sum_k}  G^0(\frac{Q}{2} + k) G^0(\frac{Q}{2} - k) 
\Gamma (Q;k,q\pp)~. 
\ee
This demonstrates that in the ladder approximation one includes scatterings 
of all orders in the particle-particle channel. Further, due to the 
independence of $\Gamma$ on internal relative momenta, the solution of 
Eq.~(\ref{Gamma_LA_eqn}) is straightforward, 
{\em viz.}
\be
\label{Gamma_LA_soln}
\Gamma (Q) ~=~\frac{-|U|}{1-|U| \chi^0(Q)}
\ee
where the pair susceptibility of the noninteracting system is equal to
\be
\label{susc0}
\chi^0(Q) = \fact \sum_q  G^0(\frac{Q}{2}+q)  G^0(\frac{Q}{2}-q)~.
\ee
Finally, using Eqs.~(\ref{pair-FT1},\ref{G2_LA_eqn},\ref{Gamma_LA_soln}) 
we can derive the pair propagator, or pair
susceptibility $\chi(Q)$, in the NSC ladder approximation:
\be
\label{pairq_LA}
{\cal G}^{pair}(Q) ~=~\chi(Q)~=~\frac{\chi^0 (Q)}{1-|U|\chi^0(Q)}
\ee

The solution of the Thouless criterion for the AHM treated in the NSC ladder 
approximation is now clear. The lifetime of a newly introduced (singlet) pair 
of electrons with crystal momentum ${\bf Q}$ will first diverge (on cooling)
when $Im \chi^0(Q)=0$ and $1-|U|\chi^0(Q)=0$. The analytical 
properties of the pair susceptibility dictate that this must occur for zero 
Matsubara frequency, namely the Thouless criterion reduces to
\be
\label{TC_eqn}
\chi^0({\bf Q},0) ~=~ \frac{1}{|U|}~.
\ee
Hereafter, the locus of points in the $T-\mu$ plane which satisfies 
Eq.~(\ref{TC_eqn}) shall be referred to as the Thouless criterion curve.
For the AHM it can be shown\cite{fukuyama91,ourlongNSCTMA} that on cooling the
transition defined by Eq.~(\ref{TC_eqn}) first occurs for $Q=0$,
and from now on we define $T_c$ in this manner.

As detailed in the introduction, we are interested in gaining an understanding 
of the effects of `feedback',\cite{nambu94} that is when the pair fluctuations that are 
included in the pair propagator are introduced into the description of the 
single-electron properties. To this end, we evaluate  the proper self-energy 
in the (same) NSC T-matrix approximation, resulting in
\be
\label{self}
\Sigma(k)=~-~\frac{|U|^2}{\beta N} \sum_Q~\chi(Q) G^0(Q-k)~.
\ee
(Note that throughout this report we ignore the Hartree term, which
can be simply absorbed into the chemical potential.)
From this quantity the fully interacting single-particle Green's function is 
found via Dyson's equation
\be
\label{dysoa_neqn}
G(k) ~=~ \frac{1}{\{G^0(k)\}^{-1}~-~\Sigma(k)}~~.
\ee
The important quantity in this paper, the single-particle density of states (DOS), 
is related to the one-electron spectral function, given by
\be
\label{1el_AKW}
A({\bf k},\omega)~=~-~\frac{1}{\pi}~Im G({\bf k},\omega + i0^+),
\ee
which leads to the DOS given by
\be
\label{1el_Nofw}
{\cal N}(\omega)~=~\frac{2}{N} \sum_{\bf k} A({\bf k},\omega).
\ee

\subsection{Partial Fractions Decomposition}

The above equations are well known. Here we present an
approach that allows for us to solve for
all of the spectral functions of finite lattices in the above-described
NSC T-matrix approximation to arbitrary numerical accuracy --- in one dimension 
we can solve for chains of lengths 
up to 128 (which has 65 inequivalent ${\bf k}$ points in the first Brillouin 
zone), while in three dimensions we have solved lattices up to a size of 
twelve cubed (which has 84 inequivalent ${\bf k}$ points in the first Brillouin
zone) --- and to then extract information about the density of states
in the thermodynamic limit. To complete the decomposition for a twelve cubed lattice at one
temperature and chemical potential can take up to ten days on a high-speed
64-bit (which is essential) alpha processor; so, it is not a simple matter 
to obtain large sets of data using this approach.

To solve for the one-particle self energy we first note that the
non-interacting pair susceptibility, defined in Eq.~(\ref{susc0}), can
be written in terms of a sum over wave vectors 
\begin{equation}
\chi^0({\bf Q},\nu_n) = \frac{1}{N} \sum_{{\bf k}} \frac{f[\xi_{\bf k}] 
+ f[\xi_{{\bf Q}-{\bf k}}] - 1}{i\nu_n - \xi_{{\bf k}} - 
\xi_{{\bf Q}-{\bf k}}}\:,
\end{equation}
where $f[\xi]$ represents the Fermi-Dirac distribution function at a 
temperature $T$ and chemical potential 
$\mu$, $\xi_{\bf k} \equiv \varepsilon_{\bf k} - \mu$, and 
$\varepsilon_{\bf k}$ is the noninteracting band electron dispersion 
({\textit e.g.}, $\varepsilon_k = -2t\cos(k)$ in one dimension).
The analytic continuation of this function
to the complex plane, denoted by $\chi \rightarrow \overline{\chi}$,
is achieved trivially  by the substitution
$i\nu_n \rightarrow z$. Thus, from Eq.~(\ref{pairq_LA}) we can write the pair 
propagator as
\be
\label{EQ:continued-NSC-pair-prop}
\bar{\chi}({\bf Q},z) = \frac{\bar{\chi}^0({\bf Q},z)}
{1-|U| \bar{\chi}^0({\bf Q},z)} 
= \frac{1/N \sum_{{\bf k}}
\frac{f[\xi_{{\bf k}}]+f[\xi_{{\bf Q}-{\bf k}}]-1}
{z - \xi_{{\bf k}} - \xi_{{\bf Q}-{\bf k}}}}
{1- |U|/N \sum_{{\bf k}}
\frac{f[\xi_{{\bf k}}]+f[\xi_{{\bf Q}-{\bf k}}]-1}
{z - \xi_{{\bf k}} - \xi_{{\bf Q}-{\bf k}}}}~.
\ee
If one simply expands the wave vector sums, it is seen that this 
function admits a partial fraction decomposition. Incorporating the requisite 
analytical properties of this 2-particle propagator (the chosen signs follow 
from the spectral representation of the 2-particle GF) we choose to write 
this expansion as 
\begin{equation} 
\label{EQ:pair-prop-partial-frac}
\bar{\chi}({\bf Q},z) =
-\frac{\textrm{sgn}(E^{(1)}_{{\bf Q}})R^{(1)}_{{\bf Q}}}{z-E^{(1)}_{{\bf Q}}} -
\frac{\textrm{sgn}(E^{(2)}_{{\bf Q}})R^{(2)}_{{\bf Q}}}{z-E^{(2)}_{{\bf Q}}} - 
\dots
\end{equation}
where the number of poles for each ${\bf Q}$ is denoted by $s_{\bf Q}$, the
energies $E_{\bf Q}^{(\ell)}$ are real, and the residues $R_{\bf Q}^{(\ell)}$ 
are strictly positive, for $\ell = 1, \dots, s_{\bf Q}$. Substituting this 
into Eq.~(\ref{self}) and completing the Matsubara frequency sum analytically 
allows us to restate the expression for the self energy in the convenient form
\begin{equation}
\bar{\Sigma}({\bf k},z) = \frac{U^2}{N}\sum_{{\bf Q}}\sum_{l=1}^{s_{{\bf Q}}}
\frac{\textrm{sgn}(E^{(l)}_{{\bf Q}})R^{(l)}_{{\bf Q}}\bigl(
f[\xi_{{\bf Q}-{\bf k}}]+b[E^{(l)}_{{\bf Q}}]\bigr)}
{z+\xi_{{\bf Q}-{\bf k}}-E^{(l)}_{{\bf Q}}}~,
\end{equation}
where $b[E]$ is the Bose-Einstein distribution for temperature $T$ with the 
chemical potential set equal to zero.

The application of the above results requires the determination of the poles 
and residues of the partial fraction expansion of the pair propagator. This is 
a complicated task, and we have completed our calculations using a computer algebra 
system allowing for a symbolic calculation that produces numerical results for the poles 
and residues.\cite{maple} Also, note that the result of this partial-fractions 
formulation is a DOS that is a series of delta functions -- we have broadened each of the 
single-particle spectral peaks with Lorentzians with a fixed broadening of 
3\% of the bandwidth, namely 0.12$t$ in one dimension, and 0.36$t$ in three dimensions.

\newpage

\section{Examination of the 1d Pseudogap - Test Case}

As an example of the application of this method, we have examined one dimensional
(1d) chains. We stress that this is not, and is not intended to be, a solution of 
the 1d AHM, which cannot be accomplished through the use of the T-matrix approximation. 
Instead we use these results as a means of demonstrating how our results
can be applied, via a finite-size scaling extrapolation to the
thermodynamic limit, to estimate the bulk value of the (pseudo)gap.
 
\begin{figure}[t]
\includegraphics[width=12.cm,height=8.5cm]{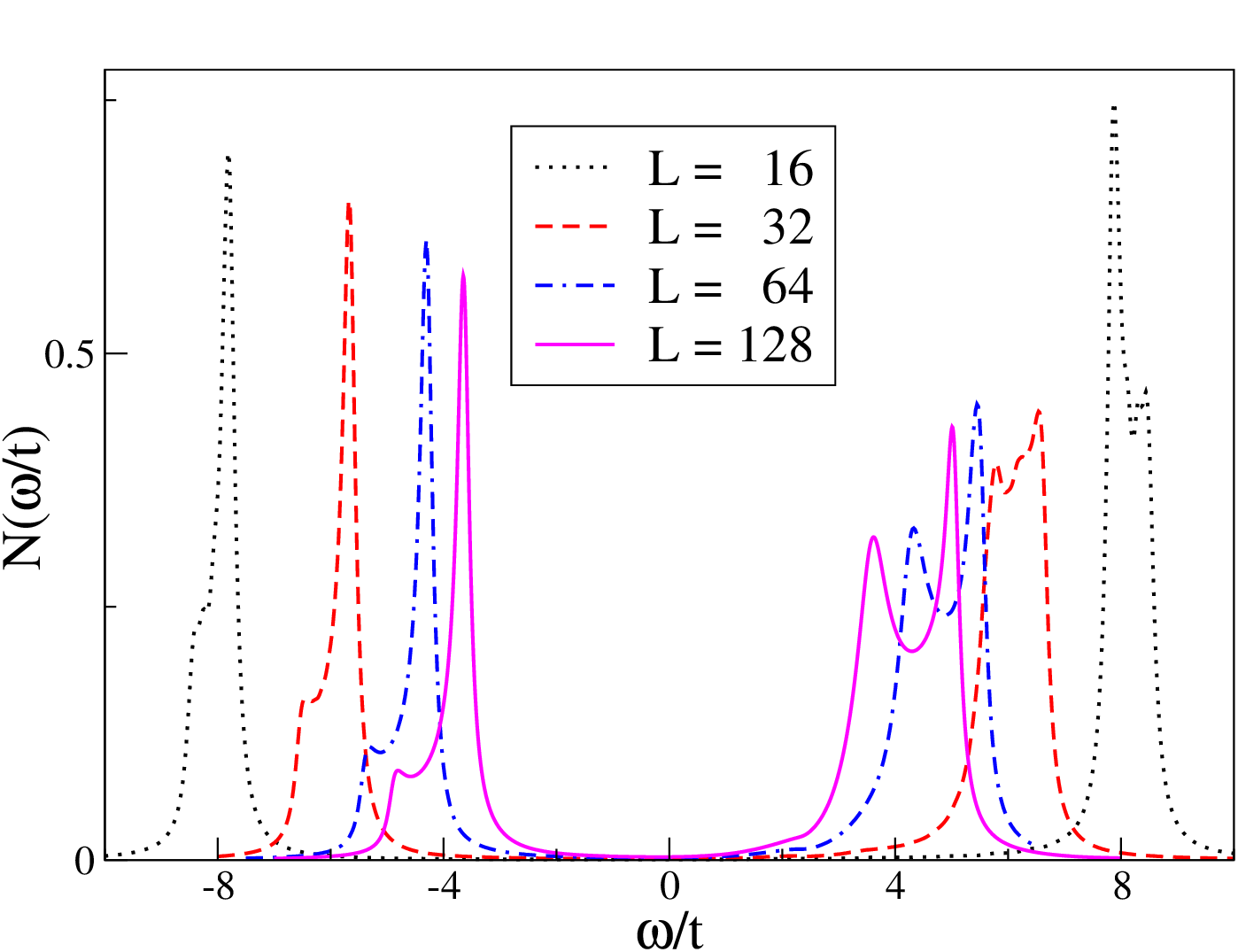}
\caption{\label{fig:DOS_1d} The density of states (DOS) of the 1d attractive Hubbard 
model evaluated in the non self-consistent T-matrix approximation for 
16$\times$1, 32$\times$1, 
64$\times$1, and 128$\times$1 chains with periodic boundary conditions, 
for $|U|/t = -2$ and a fixed (noninteracting) electronic density per lattice site of 
$\langle n \rangle = 0.5$, corresponding to 1/4 filling; 
the temperature used for each curve is 1.001~$T_c(L)$. 
Each delta-function peak of the spectral function has been broadened as a Lorentzian 
using a fixed broadening energy of 3 \% of the bandwidth --- through 
experimentation we have found that such a broadening produces smooth density 
of states curves without washing out important structure.}
\end{figure}

In Fig.~\ref{fig:DOS_1d} we show the DOS of a one-dimensional 
chain with periodic boundary conditions of varying lengths ($16\times 1$ up 
to $128\times 1$) with $|U|/t = -2$ in the non self-consistent 
T-matrix approximation for the AHM. We have taken this data for a fixed 
(non-interacting) electronic density per lattice site of $\langle n \rangle = 0.5$, 
corresponding to 1/4 filling. The temperature is fixed to be just above the BCS 
transition temperature, and to account for the fact that the BCS transition 
temperature depends on the length of the system, we have used a varying 
temperature of $1.001~T_c(L)$, thus corresponding to a fixed ratio of 
$T/T_c(L)$ for all system sizes.  

It is clear from  Fig.~\ref{fig:DOS_1d} that each DOS curve 
contains sharply peaked structures, and it is natural to associate the 
location of these peaks with something analogous to the superconducting gap. 
Of course, these data are evaluated in the normal phase, and the DOS does
not ever go strictly to zero (near $\omega/t = 0$) in the normal phase, so 
the locations of the peaks are not true superconducting gaps, but can be 
thought of as something akin to (single-particle DOS) (pseudo)gaps.

These data display strong finite size effects, and the obvious question is, 
what does the DOS look like in the thermodynamic limit? We now 
demonstrate how one can answer this question, and thus use data such as that 
shown in Fig.~\ref{fig:DOS_1d} to correctly extrapolate to the thermodynamic 
limit. This analysis will lead us to a controlled manner of both verifying
the existence of, and quantitatively estimating, the pseudogap in three dimensions.

The finite-size scaling corrections to the
wave vector sums that are encountered in this problem go as

\begin{figure}[t]
\includegraphics[width=10.5cm,height=8.5cm]{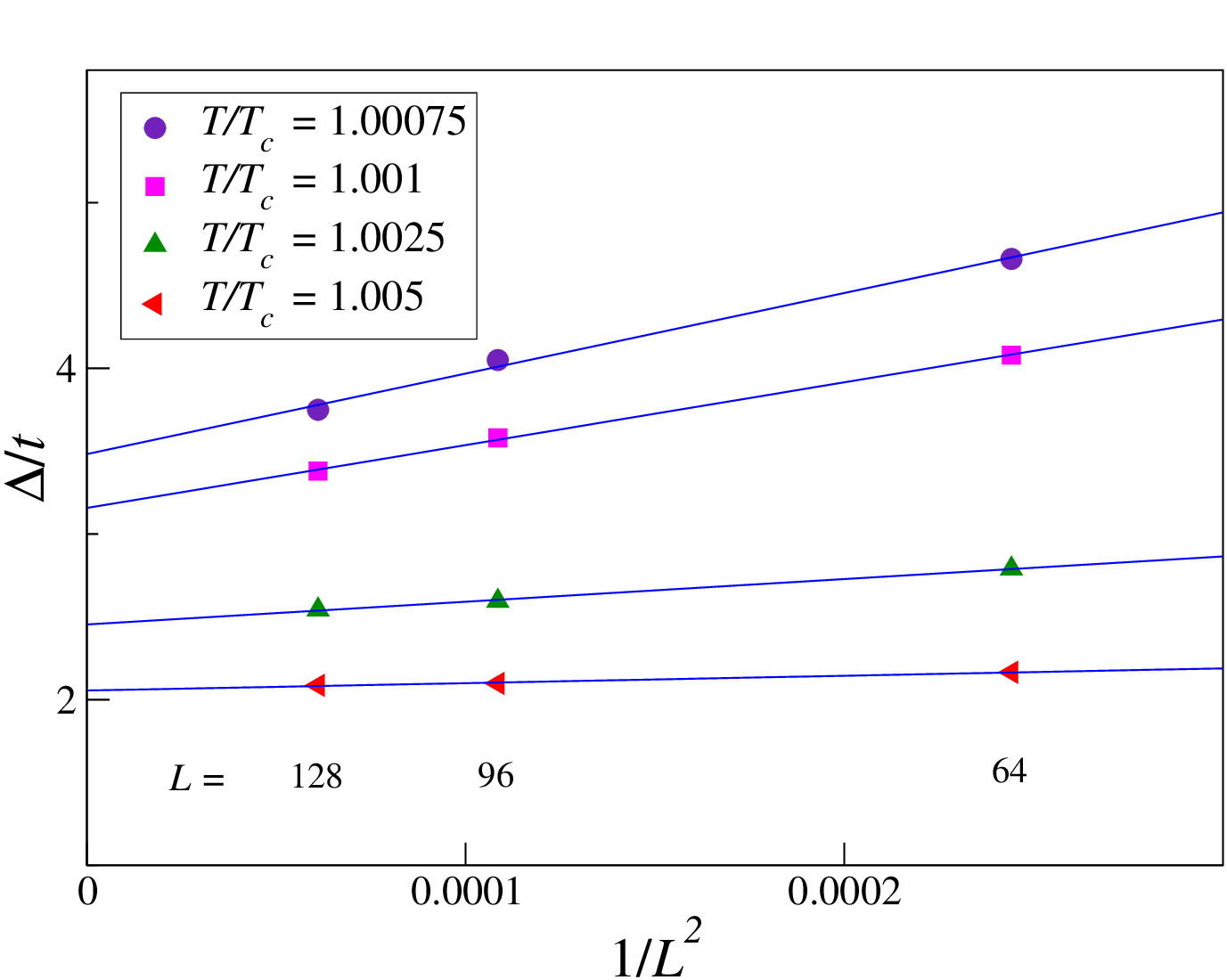}
\caption{\label{fig:DOS_1d_FSS} The locations of the peaks, labelled as $\Delta$
and expressed in units of $t$, found in the 1d DOS (\textit{e.g.}, 
see Fig.~\protect\ref{fig:DOS_1d}), plotted as a function of one over the the system's length ($L$)
squared, for a variety of temperatures (shown in the legend) just above the superconducting 
transition temperature. The same parameters as in Fig.~\protect\ref{fig:DOS_1d} are used. 
The solid lines are least squares fits to the data for each of these temperatures.}  
\end{figure}

\addvspace{0.5 truecm}

\be
\label{eq:FSS}
\frac{1}{L}\sum_q~=~\frac{1}{2\pi}~\int_{-\pi}^\pi~dq~+~\Big(\frac{1}{L}\sum_q~-~
\frac{1}{2\pi}~\int_{-\pi}^\pi~dq\Big)
~\sim~\frac{1}{2\pi}~\int_{-\pi}^\pi~dq~+~{\cal O} (\frac{1}{L^2}).
\ee 
Thus, noting that the peak locations, which from now on are
provocatively labelled as $\Delta$, from Fig.~\ref{fig:DOS_1d} 
can be associated with a strong divergence of one particular 
contribution (see below) to the self energy, 
and that the wave vector sums in Eq.~(\ref{self}) should have the same finite-size
scaling corrections as given in Eq.~(\ref{eq:FSS}), we expect that the peak locations
of, \textit{e.g.}, Fig.~\ref{fig:DOS_1d}, should approach their values in
the thermodynamic limit as $\sim 1/L^2$. This is precisely the form that we find,
as is shown in Fig.~\ref{fig:DOS_1d_FSS}.

We now focus on the values of $\Delta$ obtained from using the above-shown finite scaling
analysis extrapolated to the thermodynamic limit. To be specific, we now show that the
divergence of this (pseudo)gap as $T$ approaches $T_c$ from above is precisely what should
be found in the thermodynamic limit. This successful extraction of $\Delta(L\rightarrow\infty)$
helps to support our contention that such a finite-size scaling analysis is a reliable procedure 
for extracting the bulk value of this energy.

To this end, we first require the behaviour of the self energy as the Thouless criterion is
approached from above. It is instructive to return to the definition of the self energy (prior 
to making a partial fractions expansion) given in Eq.~(\ref{self}). All quantities 
in Eq.~(\ref{self}) are known exactly (although $\chi^0({\bf q},i\nu_n)$ requires a momentum 
sum -- see Eq.~(\ref{susc0})); nonetheless, it is not in a convenient form from which the DOS 
in the $L\rightarrow \infty$ limit can be extracted. That is, while one can successfully approximate
the wave vector integration over the first Brillouin zone with discrete momentum sums (with a 
dense mesh of $k$ points) for higher temperatures, such a method is doomed to fail for 
temperatures/chemical potentials near the Thouless curve. This is because the denominator in the 
kernel of Eq.~(\ref{self}) ({\em i.e.}, the denominator of Eq.~(\ref{pairq_LA})) approaches zero 
(for ${\bf q} =0$ and $i\nu_n =0$). To be specific, using a discrete momentum sum over the 
wavevectors of a finite lattice will result in a self energy that diverges as 
$T \rightarrow T_c$ \textit{in any dimension, for any finite lattice}. 
However, phase space arguments following the Mermin-Wagner theorem \cite{mermin66} show 
that the self energy in the form of Eq.~(\ref{self}) should {\bf not} diverge in three 
dimensions -- see below. Thus, we need to examine Eq.~(\ref{self}) more carefully to properly 
understand the distinction between one (and two) and three dimensions, and to properly extract 
the temperature dependence of the (pseudo)gap as $T_c$ is approached from above.

Since the Thouless criterion arises from the $i\nu_n = 0$ term in the Matsubara
frequency, it is prudent to treat the $n = 0$ part of the self energy sum 
separately. Then, to correctly deal with the ${\bf q} \rightarrow 0$ 
contribution, we have separated the $i\nu_n = 0$ term into two contributions 
--- the one from ${\bf q} \ne 0$ is given by
\be
\Sigma_{nzq}({\bf k}, i\omega_m) = -|U|^2\fact {\sum_{{\bf q}}}^\prime
\chi ({\bf q},0) G^0({\bf -k + q},-i\omega_m).
\label{selfnzq}
\ee       
and the one from ${\bf q}$ near zero is done by an integration (which is
correct, and as we show necessary, if we completed the entire calculation in the 
thermodynamic limit),
\be
\Sigma_{zq}({\bf k}, i\omega_m) = -|U|^2 { 1 \over \beta} \int_{-q_0}^{q_0}  
{d^dq \over (2 \pi)^d} \chi ({\bf q},0) 
{ 1 \over -i\omega_m - (\epsilon_{\bf -k + q} - \mu) }~, 
\label{selfzq}
\ee
where ${\bf q_0}$ represents an upper cutoff for the part of the first
Brillouin zone in which we will integrate over the integrand analytically, and
$d$ represents the dimensionality of the space. The prime in the summation of
Eq.~(\ref{selfnzq}) indicates that this same small ${\bf q}$ region is omitted
from $\Sigma_{nzq}$. Thus, in all expressions except
the last (Eq.~(\ref{selfzq})) we will simply perform ${\bf q}$-sums over 
finite lattices (which amounts to the trapezoidal rule for numerical integration), 
while in the latter equation it is important to integrate analytically over the integrand.

\begin{figure}
\includegraphics[width=8cm,height=8.5cm]{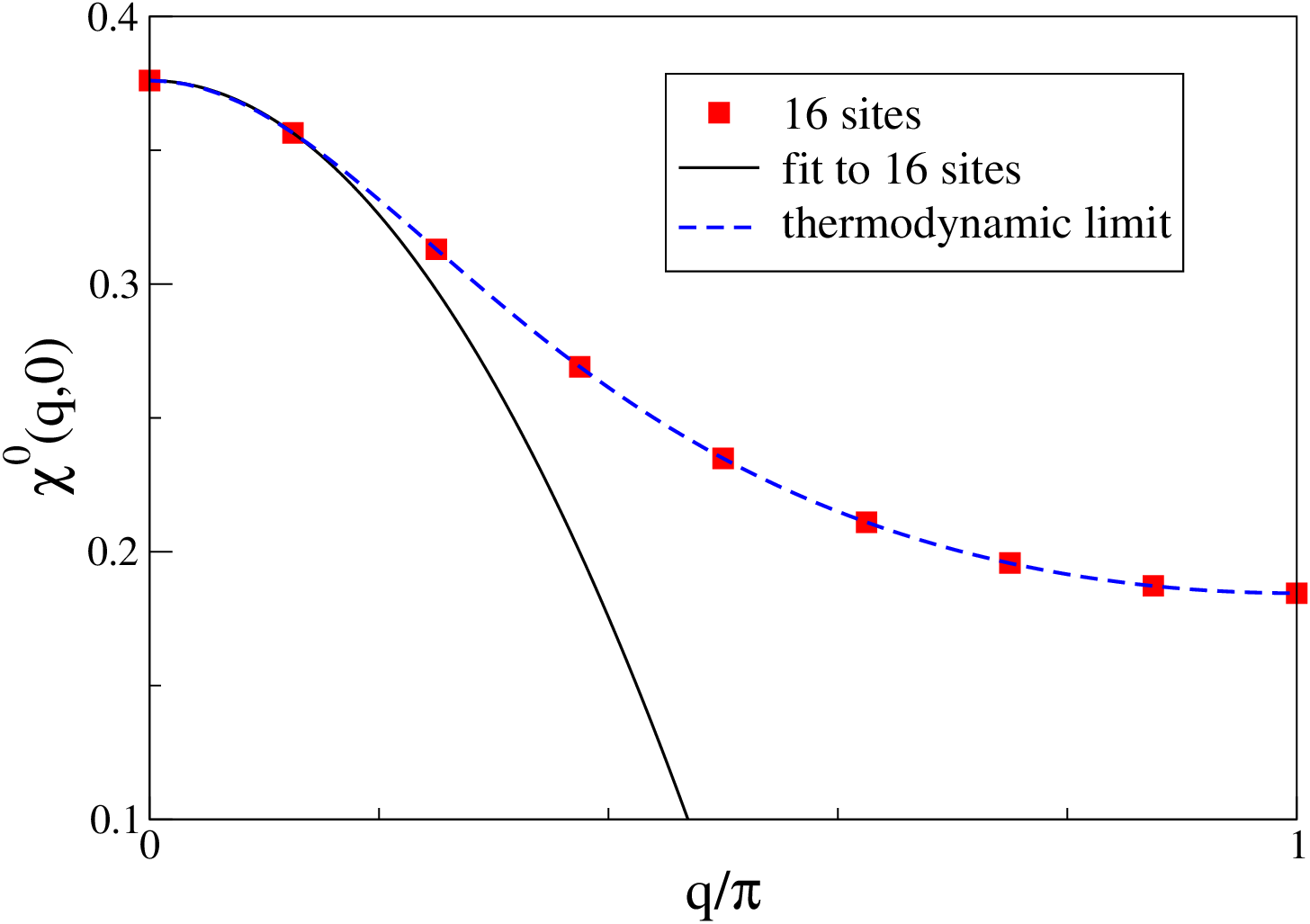}
\caption{\label{fig1_details}Real part of the non-interacting susceptibility 
{\textit vs.} $q$ in one
dimension. Red squares are for a 16 site lattice, while the green curve is based on
the fit as described in the text. The blue curve is the result in the
thermodynamic limit. Clearly, the black curve reproduces very accurately the
thermodynamic limit result at low q, even though the green curve is obtained 
on the basis of a fit to the two lowest points ($q = 0$ and $q = \pi/8$).} 
\end{figure}

To this end we require an explicit expression for $\chi^0({\bf q},0)$, which 
is difficult to obtain. Instead, we approximate it with an expression valid 
for small ${\bf q}$:
\be
\chi^0({\bf q},0) \approx a^2 - b^2 q^2,
\label{lowq}
\ee
where $q = |{\bf q}|$ and $a^2$ and $b^2$ are obtained by fits
to $\chi^0({\bf q},0)$. In one dimension this is a simple matter of fitting
the susceptibility at $q=0$ and at some small $q=q_0$, and an example case is shown in 
Fig.~\ref{fig1_details}; clearly, this is an excellent approximation for small $q$. 
Then, expanding the band dispersion $\varepsilon_{-k+q}$ for small q, the integral in 
Eq.~(\ref{selfzq}) can be written as
\begin{widetext}
\be
\label{selfZzq1}
\Sigma_{zq}(k, i\omega_m) \approx  \frac{2 |U|^2}{\pi \beta} 
~\int_0^{q_0} dq ~ \Big( \frac{a^2 - b^2 q^2}{A + Bq^2} \Big)
\Big( 1 - [ \frac{1}{2}  \epsilon_k^{\prime \prime} G^0(k,-i\omega_m)
+ (\epsilon_k^\prime)^2 (G^0(k,-i\omega_m))^2 ] q^2 \Big)~~.
\ee
\end{widetext}
We have introduced $A \equiv 1 - |U|a^2$ and $B = |U|b^2$, and note that 
both are positive numbers. Further, and most importantly, since $\chi^0(0,0)\sim 1/T$ 
at low temperatures, we obtain that $A$ is approaching zero linearly in $T$ from 
above as $T\rightarrow T_c$. That is, $A \sim |\alpha|(T-T_c)$ for alpha being 
some constant. (This clarifies the ``mean-field" label that can be applied to the 
non self-consistent T-matrix approximation.)  

All of the integrals
in this expression are elementary; the important contribution comes from
\be
\label{eq:1d_thermlimit}
\int_{0}^{q_0} dq \ \ \biggl\{ {1 \over A + Bq^2} \biggr\} = {1 \over \sqrt{AB}}
~ tan^{-1}\biggl(q_0 \sqrt{B/A} \biggr),
\ee
which diverges as $1/\sqrt{T-T_c}$ near the Thouless curve. Since we have employed
the (non self-consistent) T-matrix approximation, it is correct that
this quantity diverges in 1d; however, had we not handled this integration
analytically (and instead proceeded with lattice sums for a large but finite 
lattice) we would have obtained a divergence proportional to $1/(T-T_c)$ 
instead, which is incorrect in the thermodynamic limit.  

The behaviour of the $L\rightarrow \infty$ (pseudo)gap $\Delta$ can now be identified
through a (mean-field like) ansatz for the (pseudo)gap given by 
\be
\label{eq:self_ansatz}
\Sigma (k,i\omega_n)~=~\frac{\Delta^2}{i\omega_n + \xi_k}~~.
\ee
This ansatz has been proposed by various authors; Moreo \textit{et al.} \cite{moreo90}
remarked that the mean-field theory of Schrieffer \textit{et al.} \cite{schrieffer89}
predicts such a form.
The TPSC theory of Vilk and Tremblay \cite{vilk97} leads to the same functional form for
estimating the magnitude of the pseudogap. Also, the work of the group of Levin \cite{janko97}
has produced and critiqued this ansatz. Lastly, we note that three of the
present authors showed how such a form arises in a self-consistent T-matrix 
formulation.\cite{beach01} 

\begin{figure}[t]
\includegraphics[width=10.5cm,height=8.5cm]{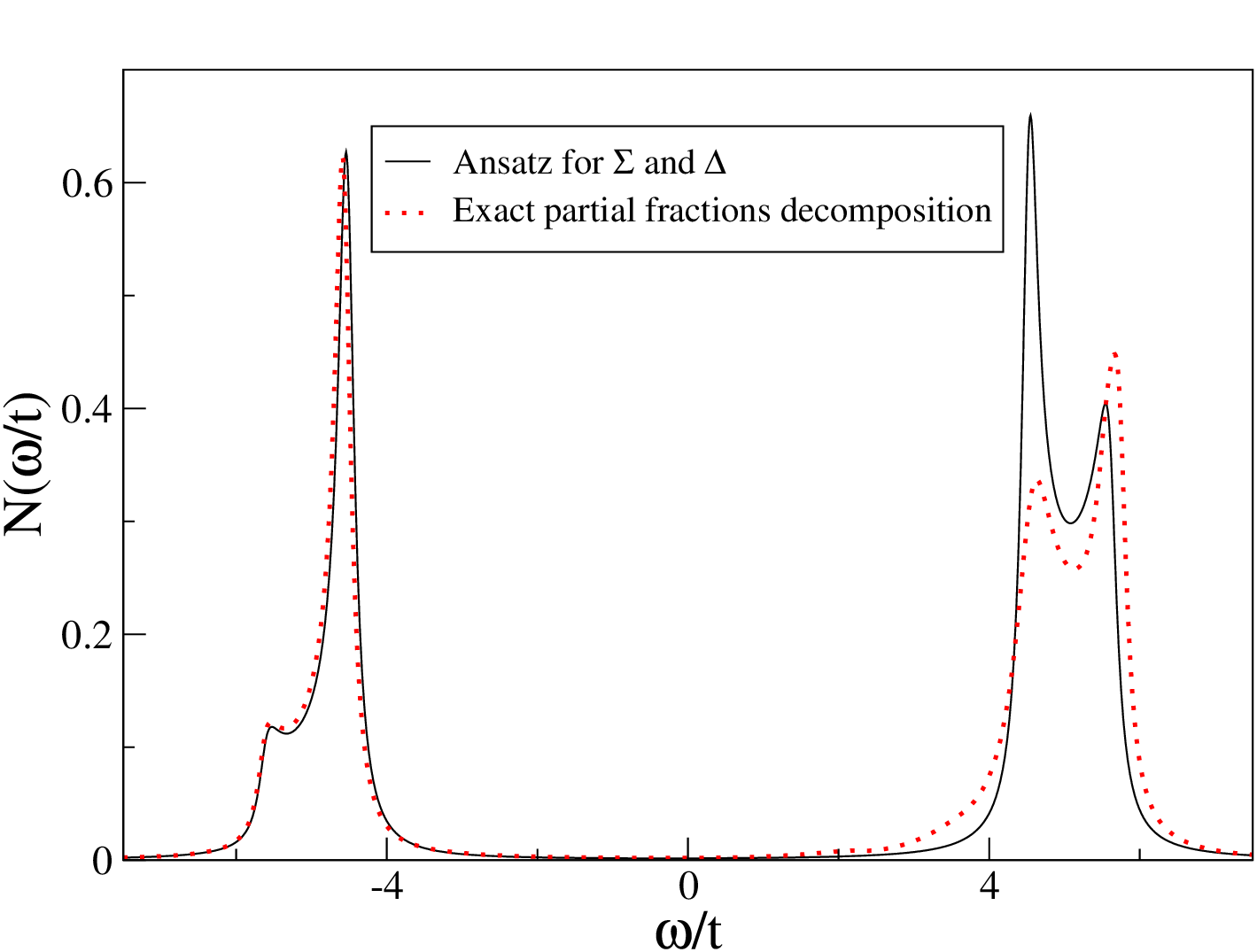}
\caption{\label{fig:1d_ansatz_comparison} A comparison of the ``exact" DOS of a finite 1d system, 
calculated within the NSC T-matrix approximation, to that predicted by 
Eq.~(\protect\ref{eq:self_ansatz}). The (pseudo)gap parameter $\Delta$ is found 
from Eq.~(\protect\ref{eq:gap_ansatz}), and then the density of states
is calculated by using a wave vector sum of the spectral functions given in Eq.~(\protect\ref{eq:Akw_ansatz}).
For these calculations the system parameters are $L$=128, $U/t = -2$, ${\langle n \rangle}_0 = 0.5$, 
and $T/T_c(L) = 1.0005$, where a broadening of 0.12$t$ has been used in both calculations.}  
\end{figure}

We augment support for this ansatz by noting that
when the (pseudo)gap parameter $\Delta$ is evaluated in the non self-consistent 
T-matrix approximation (in 1d), focussing on the $i\Omega_n$ term in the Matusbara
frequency sum, which diverges at $T_c$, one expects that
\be
\label{eq:gap_ansatz}
\Delta^2~=~\frac{|U|^2}{\beta L}~\sum_q~\frac{\chi^0(q,0)}{1-|U|\chi^0(q,0)}~~.
\ee
Here we note the following successes of this
ansatz: (I) Using the finite-size scaling result of Eq.~(\ref{eq:FSS}),
Eq.~(\ref{eq:gap_ansatz}) predicts that $\Delta^2(L) =\Delta^2(L\rightarrow\infty)
+{\cal O}(1/L^2)$, which also implies that a $1/L^2$ scaling should be found
for $\Delta$; this is precisely the form that we find, as shown in Fig.~\ref{fig:DOS_1d_FSS}.
(II) If this ansatz is used to derive the DOS when the system is close to the Thouless 
criterion line for large $L$, good agreement is found. To be specific, if one
solves for the temperature-dependent (pseudo)gap using Eq.~(\ref{eq:gap_ansatz}), and then
uses this value in Eq.~(\ref{eq:self_ansatz}), and then analytically continues to
the real axis to find the spectral function $A(k,\omega)$ through Eq.~(\ref{1el_AKW}),
one finds
\begin{widetext}
\be
\label{eq:Akw_ansatz}
A({\bf k},\omega)  
= \frac{1}{2}\Biggl(1+\frac{ \xi_{\vec{k}} }{ \sqrt{\xi_{\vec{k}}^2
+\Delta^2} }\Biggr)\delta({\omega-\sqrt{\xi_{\vec{k}}^2+\Delta^2}})
+ \frac{1}{2}\Biggl(1-\frac{ \xi_{\vec{k}} }{ \sqrt{\xi_{\vec{k}}^2
+\Delta^2} }\Biggr)\delta({\omega+\sqrt{\xi_{\vec{k}}^2+\Delta^2}})\:.
\ee
\end{widetext}
Then, to facilitate a comparison with our partial fractions results, one must broaden 
the delta functions in Eq.~(\ref{eq:Akw_ansatz}) in an analogous manner to that used in our
partial fractions decomposition, that is each delta function is replaced by a Lorentzian
with a fixed broadening of $0.12 t$. Finally, a direct comparison of the DOS from both the
exact partial fractions expansion and the above described method yields very
good agreement, in particular for the locations of the peaks closest to the
``gapped region". An example of such a comparison is shown in Fig.~\ref{fig:1d_ansatz_comparison}.
This agreement is exceptional given the crude nature of the approximation in
Eqs.~(\ref{eq:self_ansatz},\ref{eq:gap_ansatz}).

\begin{figure}[t]
\includegraphics[width=8.2cm,height=7.5cm]{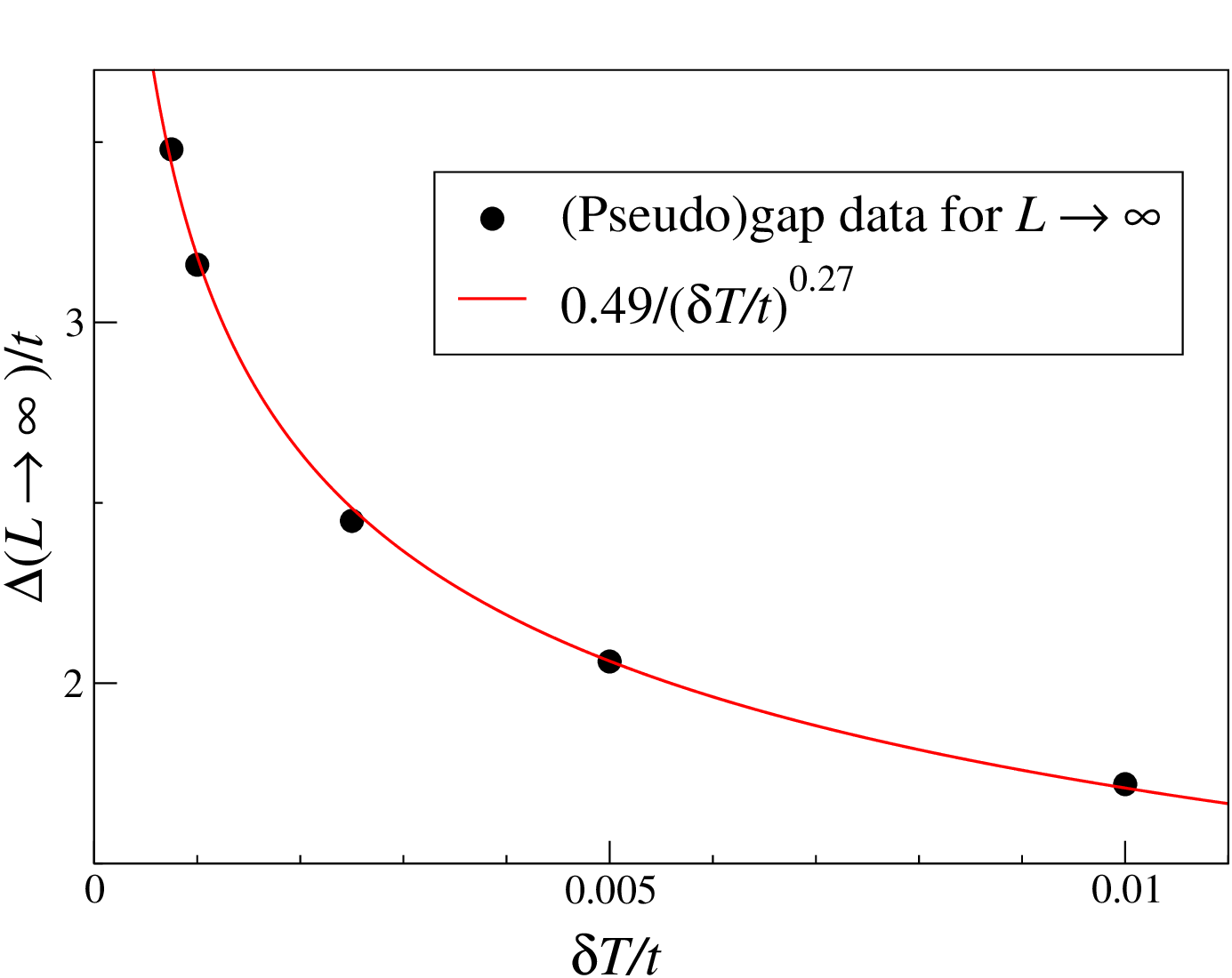}
\caption{\label{fig:1d_fig5} The locations of the peaks, extrapolated to the thermodynamic limit using
the previously described analysis, plotted as a function of the reduced temperature, defined by
$\delta T~=~(T-T_c)/T_c$, just above the superconducting transition temperature. 
The same parameters as in Fig.~\protect\ref{fig:DOS_1d} are used. The solid line shows the least-squares
fit of the data to the functional form $\Delta/t~\sim~(\delta T)^\kappa$, where we find
$\kappa~=~0.27\pm 0.03$.}  
\end{figure}

Thus, from Eq.~(\ref{eq:1d_thermlimit}) one expects that $\Sigma$ diverges as 
$1/(\sqrt{T-T_c})$, and then using Eq.~(\ref{eq:self_ansatz})
the (pseudo)gap should diverge as $1/({T-T_c})^{1/4}$ (in fact, this result
can be directly extracted from Eq.~(\ref{eq:gap_ansatz})). This is precisely the temperature
dependence that we find -- in Fig.~\ref{fig:1d_fig5} we show the $L\rightarrow\infty$
extrapolated (pseudo)gap  \textit{vs.} the reduced temperature, defined by $\delta T~=~(T-T_c)/T_c$,  
fit to the form $1/\delta T^\kappa$,
and obtain an exponent $\kappa ~=~0.27\pm 0.03$. 

Note that despite the potential difficulties of 
making a precise determination of the peak position (\textit{e.g.,} accounting for
systematic errors introduced through the use of some fixed constant Lorentzian broadened 
peaks obtained from the partial fractions decomposition), 
the obtained fit to the data is excellent, with a critical exponent very close to the 
expected value of 1/4.

We note that one may complete similar calculations in two dimensions, and again the (pseudo)gap
energy diverges to infinity as the transition temperature is approached from above. However,
in two dimensions this divergence is logarithmic, a result that follows from Eq.~(\ref{selfzq}),
and thus unless one is very close to the transition, this divergence is difficult to find. 
For example, the real-axis analysis of the T-matrix approximation in two dimensions of Metzner 
\textit{et al.} \cite{rohe01} finds spectral functions and pseudogaps similar to what we 
find, but they did 
not go to small enough reduced temperatures to observe the (unphysical) diverging pseudogap.
(What is needed to suppress this unphysical divergence, which can be traced to the
unphysical divergence of the effective vertex, is a theory beyond the non self-consistent
T-matrix approximation.\cite{beach01,vilk97})
 
We believe that the sequence of calculations outlined above for one dimension, 
\textit{viz.} (i) perform a partial fractions expansion of the pair propagator, 
thus obtaining the poles and residues of
the one-particle Green's function (for a finite lattice) to arbitrary numerical
accuracy, and (ii) extrapolating to
the thermodynamic value of the (pseudo)gap as a function of the reduced temperature
using Eq.~(\ref{eq:FSS}), is justified, in part by the above demonstration. 

\section{A Pseudogap in Three Dimensions}

We now describe the result that is central to the goal and message of this paper,
\textit{viz.} a single-particle DOS pseudogap above the superconducting $T_c$
can be present in three dimensions, even in the weak coupling limit. 

To begin, we first note the qualitative change that arises from an analysis
of the behaviour of the self energy in three \textit{vs.} one (or two) dimensions. 
That is, according to Eq.~(\ref{selfzq}), even when the denominator of $\chi(q,0)$ 
vanishes at $q=0$ the self
energy remains finite. It is this difference that makes the study of the AHM
on a  three-dimensional lattice so attractive (no pun intended). That is, one 
correctly predicts a true phase transition at a nonzero temperature, and the 
Mermin-Wagner theorem does not disallow such a transition in three dimensions -- it 
is the true superconducting phase transition that such a system will undergo. 

We have repeated the analysis of the previous section for a three-dimensional
simple cubic lattice with periodic boundary conditions. As system parameters
we have used $|U|/t = 6$ and 3 (that is, $|U|=W/2$ and $|U|=W/4$, where
$W=12t$ is the noninteracting bandwidth in three dimensions). 
The first of these ratios is the same as the on-site attractive interaction 
energy to bandwidth that was used in the previous section's study of the 1d AHM, while
the second is in the weak-coupling limit. 
Also, in part due to the interest in the cuprate superconductors and doped Mott insulators, 
we have examined a (noninteracting) electronic density of 
$\langle n \rangle_0~=~1$, \textit{viz.} half filling. 

\begin{figure}[t]
\includegraphics[width=12.cm,height=8.5cm]{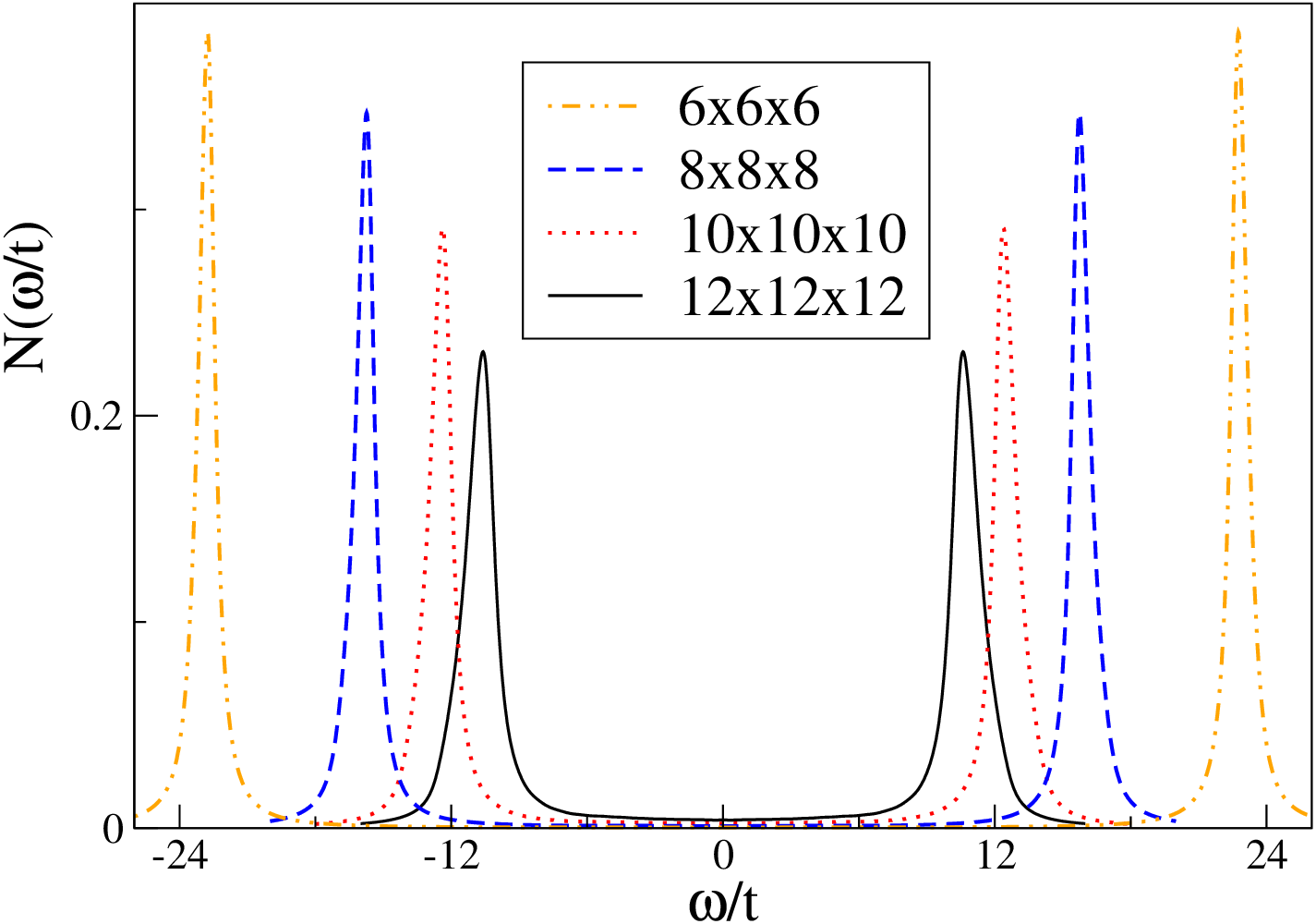}
\caption{\label{fig:3d_DOS} The single-particle DOS
for simple cubic lattices of size $6^3$,~$8^3$,~$10^3$~, 
and $12^3$, for $|U|/t = 6$ and 1/2 filling, at a 
temperature of $T/T_c(L) = 1.0001$. We have employed a fixed
broadening of 3\% of the noninteracting bandwidth, namely 0.36$t$.}
\end{figure}

Results for the DOS for different size lattices are shown in Fig.~\ref{fig:3d_DOS}
for $|U|/t = 6$, for a reduced temperature of 0.0001.
As in the previous section, we have broadened the delta functions that emerge from the
partial fractions decomposition with a fixed energy equal to 3 \% of the noninteracting
bandwidth, which is 0.36$t$ for the simple cubic lattice. 
It is to be emphasized that there are \textit{no} true superconducting gaps in these spectra,
in that the DOS is never reduced to zero at the Fermi level (which is located at $\omega = 0$).

\begin{figure}[t]
\includegraphics[width=12.cm,height=8.5cm]{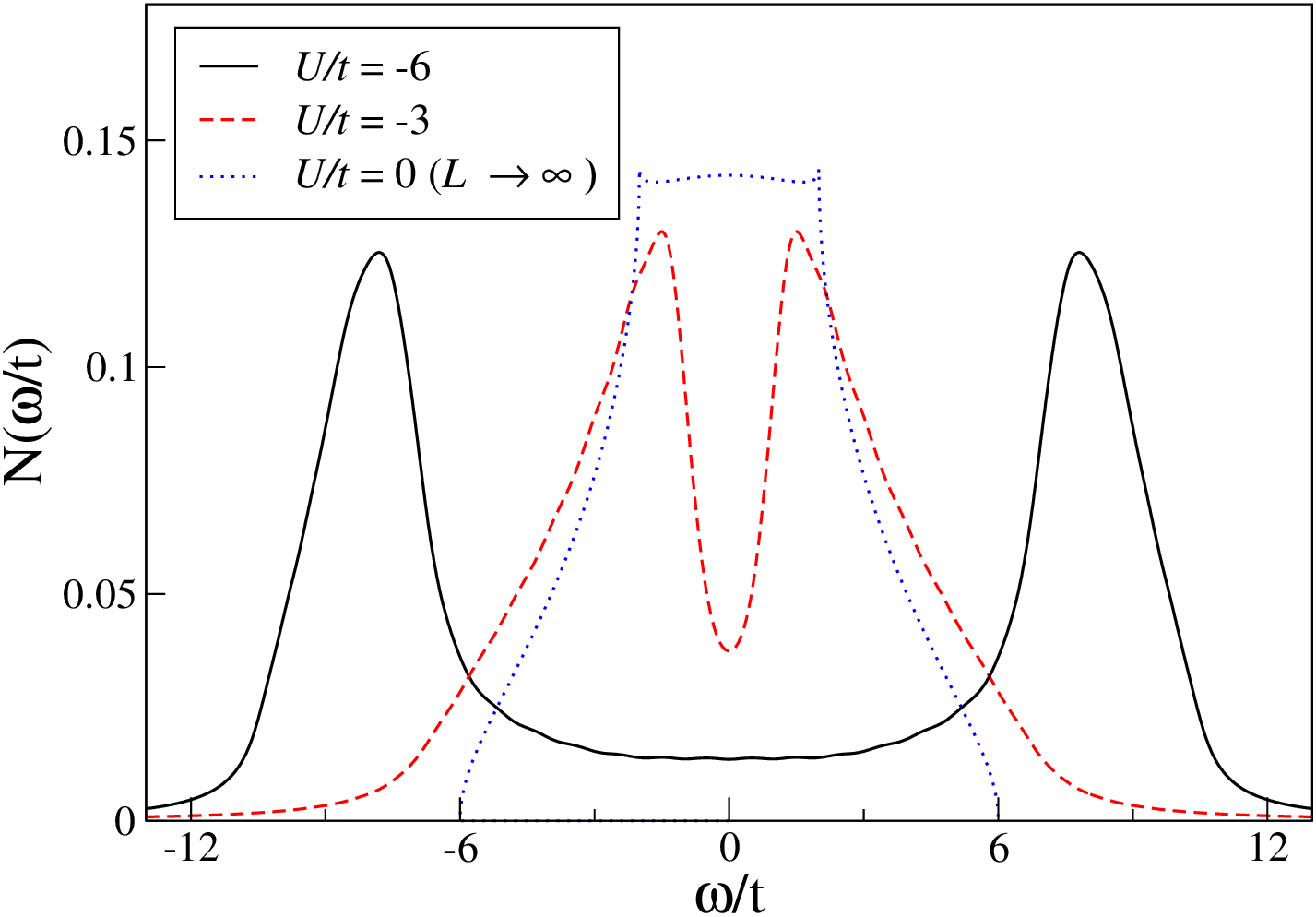}
\caption{\label{fig:3d_differingUs} DOS for $|U|/t$ = 6 and 3 at half filling, for a $12^3$ simple 
cubic lattice, at a temperature of $T~=~1.001~T_c$. In this plot we have used the same broadening 
(0.36t) for both $U \not=$ 0 curves. As a reference curve we also show the DOS of the noninteracting
system in the thermodynamic limit.}
\end{figure}

How does the DOS develop as a function of the attractive Hubbard energy $U$? 
In Fig.~\ref{fig:3d_differingUs} we show the change of the DOS as $|U|/t$ is decreased 
from 6 to 3, using a reduced temperature of $0.001$ for each value of $|U|$,
for a $12^3$ lattice; that is, $T/T_c(|U|,L)~=~1.001$. As a reference curve, we also 
show the $U=0$ noninteracting DOS in the thermodynamic limit. 

\begin{figure}[t]
\includegraphics[scale=0.5,bb=100 0 600 600]{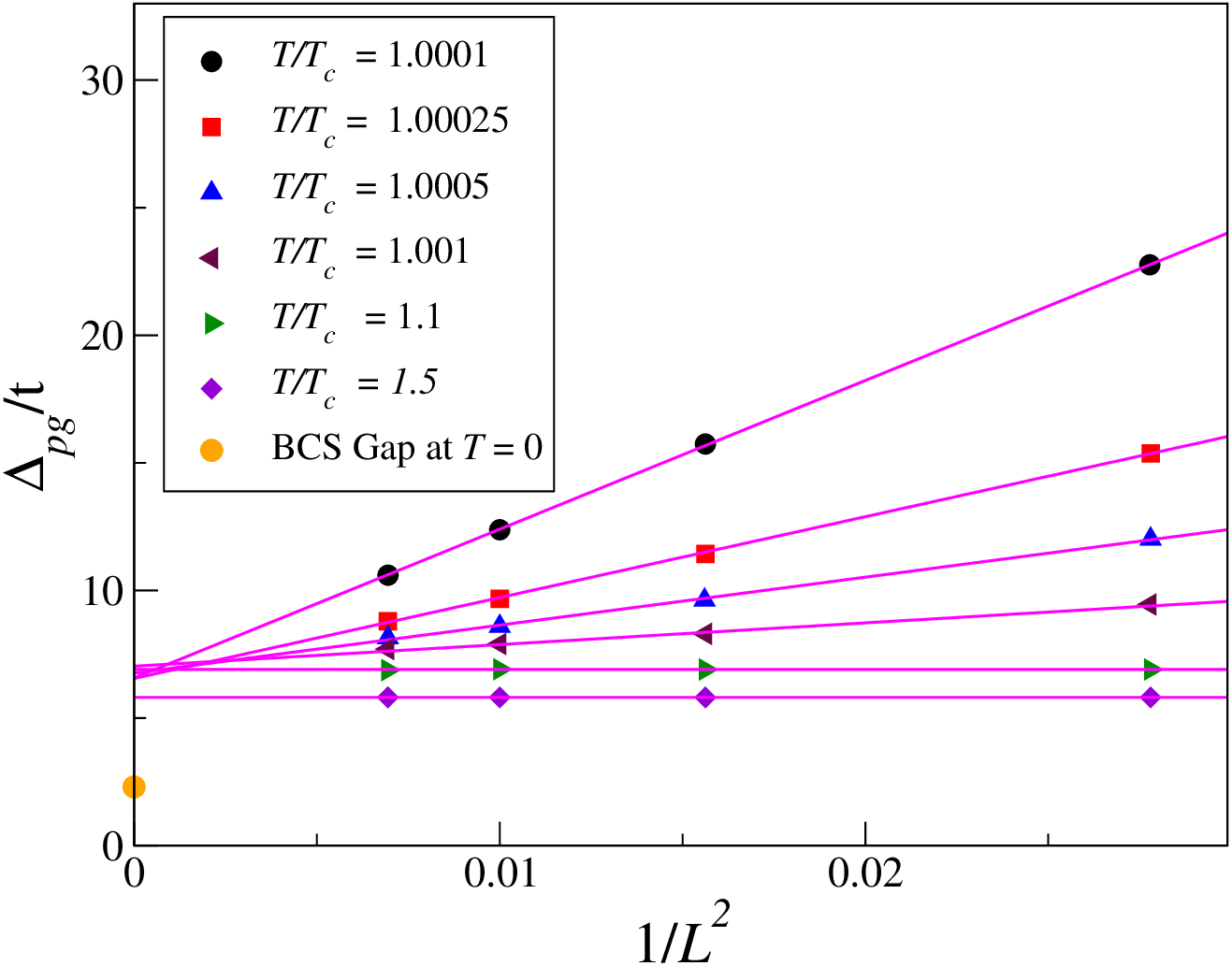}
\caption{\label{fig:3d_gapscaling_Um6} The peak locations of the DOS, labelled as $\Delta_{pg}$
and expressed in units of $t$, for 3d simple cubic lattices of size $6^3$,~$8^3$,~$10^3$~,
and $12^3$, for $|U|/t = 6$ and 1/2 filling,
plotted as a function of one over the the system's linear size ($L$)
squared, for a variety of temperatures (shown in the legend) just above the superconducting
transition temperature. The solid lines are least squares fits for each temperature. The
$T=0$ superconducting gap (in the thermodynamic limit), found from the solution of the BCS 
gap equation, is also shown.}
\end{figure}

The scaling of the peak positions, which we now label
as $\Delta_{pg}$, \textit{vs.} system size is shown in 
Fig.~\ref{fig:3d_gapscaling_Um6},
and we again find a finite-size scaling behaviour that clearly goes as $1/L^2$. From this figure 
one sees that all extrapolated values for small reduced temperatures are very close to one another 
(approximately 6.5$t$ for these material parameters), and do not go to zero, or infinity, as the 
reduced temperature is lowered towards zero. Further, one can use BCS theory\cite{bardeen57} 
to calculate the superconducting energy gap at $T=0$, and for 1/2 filling and $|U|/t = 6$ one 
finds a gap of about 2.3$t$, roughly 1/3 of the energy of the pseudogap found in 
Fig.~\ref{fig:3d_gapscaling_Um6} (which is evaluated just above $T_c$, and not $T=0$!).

How does the pseudogap behave as a function of temperature? As seen in 
Fig.~\ref{fig:3d_gapscaling_Um6},
for $\delta T = 0.1$ the pseudogap is roughly the same as its value at lower temperatures,
in the thermodynamic limit. In
fact, for a temperature of 1.5$T_c$ we find a pseudogap that is about 85\% of its value as
$\delta T \rightarrow 0$. So, the pseudogap energy does not change appreciably with temperature
for small reduced temperatures (less than 0.1). At sufficiently high temperatures the 
pseudogap disappears, but we find that it does so by the DOS filling in, as opposed 
to the pseudogap energy going to zero.

\begin{figure}[t]
\includegraphics[width=12.cm,height=8.5cm]{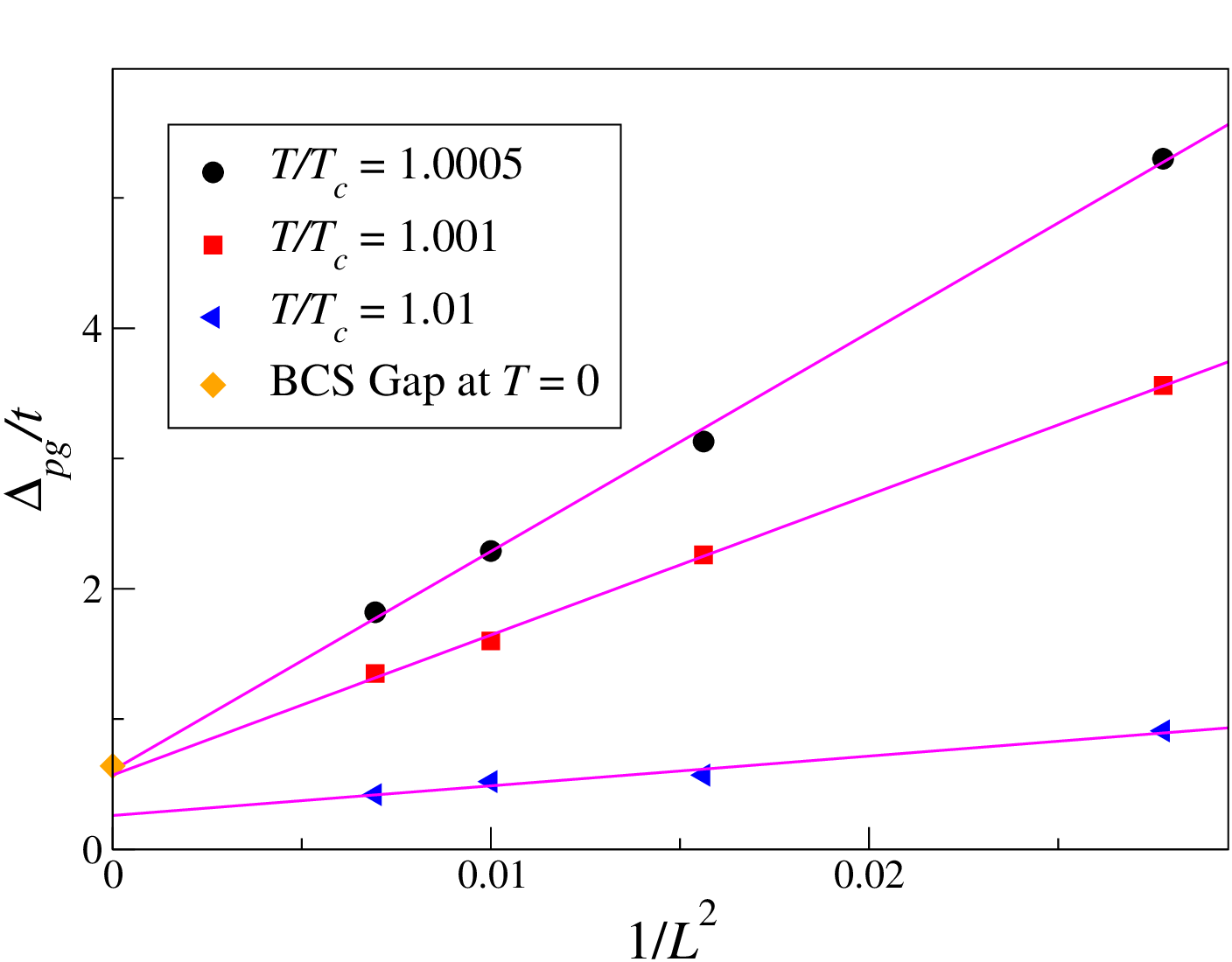}
\caption{\label{fig:3d_gapscaling_Um3}  The same quantities as in 
Fig.~\protect\ref{fig:3d_gapscaling_Um6} are plotted 
for $|U|/t = 3$. Note that the extrapolated values of the pseudogap, at temperatures 
just above the superconducting transition temperature, are within 5\% of the $T=0$ BCS gap.}
\end{figure}

As to whether or not the pseudogap persists for lower couplings, we have completed a similar
analysis for $|U|/t = 3$ at half filling, and again find evidence for a pseudogap --
the analogue of Fig.~\ref{fig:3d_gapscaling_Um6} is shown in 
Fig.~\ref{fig:3d_gapscaling_Um3}.\cite{smallerdelta}
We find the striking result that the extrapolated gap for smaller reduced temperatures 
is within 5\% of the BCS gap! We emphasize that this matching of energy ``gaps" is found 
between a $T=0$ BCS gap and a $T_c < T {< \atop \sim} < 1.001 T_c$ pseudogap.

Our result, that is the existence of a pseudogap in a model of a three-dimensional strongly
correlated electronic system, agrees with the earlier (three-dimensional) work of 
Levin \textit{et al.}.\cite{janko97} However, they propose that the pseudogap they find can 
be understood in terms of a T-matrix resonance caused by intermediate to large $U$, whereas 
we are finding the full value of the BCS gap above $T_c$ even for weak coupling. 
 
For smaller reduced temperatures and large lattices our partial fractions algorithm fails,
due to the high accuracy that must be enforced to find converged pole locations and residues.
To be specific, we were unable to find a converged partial fractions decomposition even
after 30 days of computation for $L=12$ for reduced temperatures less than 0.0005.

Clearly, a thorough search of parameter space, that is for other electronic densities
and for other ratios of $|U|/t$, is called for. Unfortunately, as mentioned earlier, it takes
roughly 10 days to get each data point for a 12$^3$ lattice (for one temperature and one chemical
potential), so new implementations of this approach are being explored in the hope of 
finding a more expedient algorithm. 

\section{Conclusions:}

We have introduced a technique that allows for the extraction of the \textit {real-axis 
dynamical properties} of the pair susceptibility, the self energy, and the spectral 
functions, for a strongly correlated electronic system when unrenormalized propagators 
are used. We have applied this technique to the attractive Hubbard model, and examined 
the properties of the density of states in the thermodynamic limit. Of particular interest
for us is three dimensions, since a true transition at a nonzero temperature to a superconducting
phase should be present for this model of correlated electrons, and for weak coupling, 
described quite well by the non self-consistent T-matrix approximation. We find a pseudogap 
in three dimensions that persists in the weak coupling limit, and whose energy is 
comparable to the $T=0$ BCS gap energy in the same weak coupling limit; for larger 
$|U|/t$ we find a pseudogap energy that is much larger than the $T=0$ BCS gap energy  .
  
\acknowledgments

We wish to thank Andrew Callan-Jones and Andr\'e-Marie Tremblay for useful discussions.
This work was supported, in part, by the NSERC of Canada. One of us (FM) acknowledges the
support of the Quantum Materials Program of the CIAR, and ICORE
(Informatics Circle of Research Excellence) of Alberta.


\begin{thebibliography} {999}
\bibitem{KT} 
We make this statement with the understanding that we are not attempting to
discuss Kosterlitz-Thouless physics in this paper.
\bibitem{smallerdelta} We note that to complete the $|U|/t = 3$ analysis we
have had to use a smaller broadening parameter of 0.12$t$, in contrast to
the value of 0.36$t$ used for $|U|/t = 6$. This is because the energy
scales are much smaller, and a broadening of 0.36$t$ is in fact more than 50\% of
the BCS gap; to properly resolve the energies for smaller $|U|/t$ we
thus used a smaller broadening.
\end{thebibliography}
\end{document}